\documentclass[a4paper]{article}
\usepackage{amsmath,graphicx}

\usepackage{pgfplots}
\usepackage{import}

\usepackage{amssymb}
\usepackage{siunitx}
\usepackage{cleveref}
\usepackage{tabularx, etoolbox, booktabs}
\usepackage{multirow} 

\usepackage{bbding}
\usepackage{microtype}
\usepackage{pifont}
\usepackage{balance}

\usepackage{tikz}
\usepackage{standalone}
\usepackage{pgfplots}

\usetikzlibrary{positioning,chains,calc,shapes.geometric, shadows, shapes.misc, fit, arrows, shapes.callouts}
\usepgfplotslibrary{groupplots}
\usetikzlibrary{backgrounds}

\definecolor{yellow}{RGB}{255, 198, 0}
\definecolor{orange}{RGB}{255, 130, 0}
\definecolor{blue}{RGB}{0, 32, 91}
\definecolor{red}{RGB}{198, 53, 39}
\definecolor{magenta}{RGB}{138, 27, 97}
\definecolor{lightblue}{RGB}{0, 159, 223}
\definecolor{green}{RGB}{0, 155,119}
\definecolor{lightgreen}{RGB}{132, 189,0}
\definecolor{greenyellow}{RGB}{208, 223,0}

\usepackage{spconf}
\usepackage[acronym, toc, nonumberlist]{glossaries}
\newacronym{AM}{AM}{acoustic model}
\newacronym{ASR}{ASR}{automatic speech recognition}
\newacronym{ATF}{ATF}{acoustic transfer function}
\newacronym{AUC}{AUC}{area under the curve}
\newacronym{BAN}{BAN}{blind analytic normalization}
\newacronym{BLSTM}{BLSTM}{bidirectional long short term memory network}
\newacronym{BSS}{BSS}{blind source separation}
\newacronym{CE}{CE}{cross entropy}
\newacronym{CNN}{CNN}{convolutional neural network}
\newacronym{CSBE}{CSBE}{combined sub-band energy}
\newacronym{DAN}{DAN}{deep attractor network}
\newacronym{DCF}{DCF}{decision cost function}
\newacronym{DC}{DC}{deep clustering}
\newacronym{DFT}{DFT}{discrete Fourier transformation}
\newacronym{DNN}{DNN}{deep neural network}
\newacronym{DOA}{DoA}{direction of arrival}
\newacronym{EEG}{EEG}{electroencephalography}
\newacronym{EER}{EER}{equal error rate}
\newacronym{EM}{EM}{expectation maximization}
\newacronym{FF}{FF}{feed-forward}
\newacronym{FPR}{FPR}{false positive rate}
\newacronym{G-cACGMM}{G-cACGMM}{Gaussian complex angular central Gaussian mixture model}
\newacronym{GCC}{GCC}{generalized cross-correlation}
\newacronym{GEV}{GEV}{generalized eigenvalue}
\newacronym{GMM}{GMM}{Gaussian mixture model}
\newacronym{GRU}{GRU}{gated recurrent unit}
\newacronym{HMM}{HMM}{hidden Markov model}
\newacronym{IBM}{IBM}{ideal binary mask}
\newacronym{ICA}{ICA}{independent component analysis}
\newacronym{ILD}{ILD}{intra-channel level difference}
\newacronym{IPD}{IPD}{intra-channel phase difference}
\newacronym{IRM}{IRM}{ideal ratio mask}
\newacronym{LCMV}{LCMV}{linearly constrained minimum variance}
\newacronym{LPC}{LPC}{linear predictive coding}
\newacronym{MM}{MM}{majorize-minimization or minorize-maximization}
\newacronym{MSE}{MSE}{mean squared error}
\newacronym{MS}{MS}{miimum statistics}
\newacronym{MVDR}{MVDR}{minimum variance distortionless response}
\newacronym{NMF}{NMF}{non-negative matrix factorization}
\newacronym{PA}{PA}{permutation alignment}
\newacronym{PCA}{PCA}{principal component analysis}
\newacronym{PDF}{PDF}{probability density function}
\newacronym{PESQ}{PESQ}{perceptual evaluation of speech quality}
\newacronym{PIT}{PIT}{permutation invariant training}
\newacronym{PSD}{PSD}{power spectral density}
\newacronym{RIR}{RIR}{room impulse response}
\newacronym{RNN}{RNN}{recursive neural network}
\newacronym{ROC}{ROC}{receiver operating characteristic}
\newacronym{RSAN}{RSAN}{recurrent selective attention network}
\newacronym{RTF}{RTF}{relative transfer function}
\newacronym{SAD}{SAD}{speech activity detection}
\newacronym{SDR}{SDR}{signal to distortion ratio}
\newacronym{SIR}{SIR}{signal to interference ratio}
\newacronym{SNR}{SNR}{signal to noise ratio}
\newacronym{STD}{STD}{standard deviation}
\newacronym{STFT}{STFT}{short time Fourier transform}
\newacronym{TDOA}{TDoA}{time difference of arrival}
\newacronym{TPR}{TPR}{true positive rate}
\newacronym{TV-cGMM}{TV-cGMM}{time-variant complex Gaussian mixture model}
\newacronym{VAD}{VAD}{speech activity detection}
\newacronym{VEM}{VEM}{variational expectation maximization}
\newacronym{WER}{WER}{word error rate}
\newacronym{WRN}{WRN}{wide residual network}
\newacronym{WSJ}{WSJ}{Wall Street Journal}
\newacronym{cACGMM}{cACGMM}{complex angular central Gaussian mixture model}
\newacronym{cBMM}{cBMM}{complex Bingham mixture model}
\newacronym{cWMM}{cWMM}{complex Watson mixture model}
\newacronym{vMF-cACGMM}{vMF-cACGMM}{von-Mises-Fisher complex angular central Gaussian mixture model}
\newacronym{vMFMM}{vMFMM}{von-Mises-Fisher mixture model}
\newacronym{vMF}{vMF}{von-Mises-Fisher}

\title{\vspace{-0.5cm}Statistical and Neural Network Based Speech Activity Detection \\  in Non-Stationary Acoustic Environments}
\name{Jens Heitkaemper, Joerg Schmalenstroeer, Reinhold Haeb-Umbach}
\address{Paderborn University, Department of Communications Engineering, Paderborn, Germany \\ \small \tt \{heitkaemper, schmalen, haeb\}@nt.upb.de}

\begin{document}
\ninept
\maketitle
\begin{abstract}
Speech activity detection (SAD), which often rests on the fact that the noise is ``more'' stationary than speech, is particularly challenging in non-stationary environments, because the time variance of the acoustic scene makes it difficult to discriminate  speech from noise. We propose two approaches to SAD, where one is based on statistical signal processing, while the other utilizes neural networks. The former employs sophisticated signal processing to track the noise and speech energies and is meant to support the case for a resource efficient, unsupervised signal processing approach.
The latter introduces a recurrent network layer that operates on short segments of the input speech to do temporal smoothing in the presence of non-stationary noise. The systems are tested on the Fearless Steps challenge database, which consists of the transmission data from the Apollo-11 space mission.
The statistical SAD  achieves comparable detection performance to earlier proposed neural network based SADs, while the neural network based approach leads to a decision cost function of $\SI{1.07}{\percent}$ on the evaluation set of the 2020 Fearless Steps Challenge, which sets a new state of the art.

\end{abstract}

\section{Introduction}

\Gls{SAD} is an integral part of many speech processing pipelines. For example, it is used to define speech on/offsets for diarization \cite{Anguera12Diarization,Mertens12Diarization}, to reduce  the computational effort of speech recognition systems by specifying the temporal regions for \gls{ASR} decoding \cite{Dumpala17VadInAsr,Ryant2013SpeechAD}, or to support noise power estimation in speech enhancement algorithms \cite{Martin18VadBook}.
Indeed, \gls{SAD} has been the focus of research efforts for years \cite{Sohn99VAD,Gerkmann08Vad,Hughes13RnnVad,Liu19VAD}.

Traditionally, \gls{SAD} is formulated as a statistical hypothesis test employing probabilistic models, such as Gaussians, mixtures of Gaussians, or Laplacian distributions \cite{Sohn99VAD,Chang06StatVad,Graciarena13SAD,Ziaei15VadFearless}.
During the last decade, however, \glspl{DNN} have achieved impressive results on some of the more taxing \gls{SAD} tasks, outperforming the traditional approaches \cite{Hughes13RnnVad,Saon13IbmVAD,Vafeidias19CnnFearless}. Here, \gls{SAD} is formulated as a supervised learning problem by presenting  the speech signal at the network's input and the  class labels (speech / no speech) as training targets at the output.

One of these challenging tasks is the \gls{SAD} on the transmission data from the Apollo-11 mission which is one of the objectives of  the Fearless Steps challenge \cite{Hansen18Fearless}. 
The signals are degraded due to high channel noise, system noise, attenuated signal bandwidth, analog tape ageing, etc..
Furthermore, the noise conditions and signal-to-noise ratios change rapidly over time and channels.
Many characteristics of these signals can also be observed in analog speech transmission over High Frequency (HF) radio bands.

In the 2019 edition of the challenge, the top performing neural network-based system achieved a $\SI{66.4}{\percent}$ improvement over the baseline system \cite{Vafeidias19CnnFearless}. The latter consisted of two \glspl{GMM} applied to the one-dimensional \gls{PCA} of a concatenation of multiple high dimensional noise robust features \cite{Ziaei15VadFearless}. The neural network, on the contrary, consisted of several \gls{CNN} layers with subsequent \gls{RNN} layers to exploit temporal information, and employed majority voting on the output of multiple networks.
Additionally, a post filter was used to smooth unwanted oscillations in the network's decisions over time.

One should note, however, that the baseline GMM system is rather simplistic. 
If one used more sophisticated signal processing techniques one should be able to come closer to the performance of the neural network while still requiring considerably less computational and memory resources. To show that this is indeed possible, we present a statistical SAD, which combines multi-layer minimum statistics-based noise estimation and Wiener filter-based enhancement, followed by an energy-based \gls{SAD}.
We show that the presented statistical \gls{SAD} achieves competitive results compared to those published previously on the Fearless Steps dataset \cite{Kaushik18FearlessCNN,Sharma19FearlessStat}, thereby closing the gap between statistical \cite{Ziaei15VadFearless,Kaushik18FearlessCNN} and neural network based \gls{SAD} \cite{Vafeidias19CnnFearless,Sharma19FearlessStat}.

Additionally, we propose an improved neural network based \gls{SAD} which achieves even better results.
It is a \gls{CNN}-based system inspired by the latest advances in sound event detection \cite{Ebbers2019WeaklySAD}. The network topology is similar to the one in \cite{Vafeidias19CnnFearless}. But unlike that system, we introduce a segment \gls{RNN} which conducts temporal smoothing inside the network rather than by a postfilter.
The segment \gls{RNN} operates on a fixed segmentation of the input signal to control the context observed by the \gls{RNN}.
This is different from other layer types like the hierarchical multiscale \gls{RNN} \cite{Chung16HMLSTM} and the \gls{RNN}-based approach presented in \cite{Kong15SegmentalRNN}, which have to learn the segmentation in addition to the segment labeling.

Furthermore, the network calculates multiple predictions per time frame, which are subsequently aggregated.
However, instead of using different temporal context lengths as in \cite{Zhang16ContextDnnSAD}, we aggregate information from different segments with overlapping input frames after the \gls{RNN} layer for an automatic smoothing.
We show that the segment \gls{RNN} outperforms all previously published results on the fearless dataset w.r.t. the \gls{DCF} measure.


The remainder of this paper is structured as follows.
In Section 2 the statistical \gls{SAD} system is described, and in Section 3 the neural network-based \gls{SAD} is introduced. Section 4 includes an evaluation of the two systems on the Fearless Steps challenge dataset \cite{Hansen18Fearless}.


\section{Statistical SAD}\label{sec:stat_sad}
The proposed statistical \gls{SAD} is following the idea of \cite{Warsitz2007} to conduct a  two-stage processing: In the first stage denoising is carried out, and in the second stage a time domain energy criterion is applied to decide on speech activity.
Although this \gls{SAD} works reliably for signals with medium ($\geq \SI{5}{dB}$) \glspl{SNR} and stationary noise conditions, it fails for the highly non-stationary noise of the Fearless dataset for multiple reasons:
The recordings exhibit a large variety of different and changing noise types, and they have much lower \gls{SNR} values.
Additionally, the signal magnitude of the active speakers varies  by several orders of magnitude, even in the same recording, making it difficult to decide whether the noise floor or a speaker with low energy signals is observed.

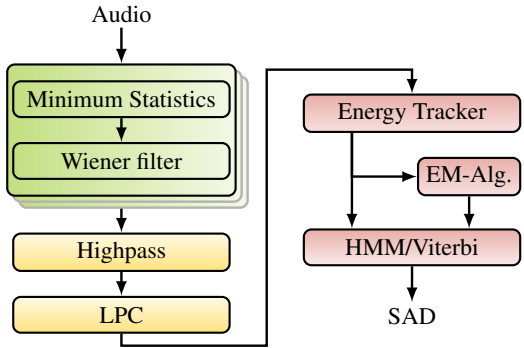
\begin{figure}[t]
	\centering
	\tikzset{%
  block/.style    = {draw, thick, rectangle, minimum height = 1.5em, minimum width = 9em, fill=white, align=center, rounded corners=0.1cm},
  sum/.style      = {draw, circle, node distance = 2cm}, 
  cross/.style={path picture={\draw[black](path picture bounding box.south east) -- (path picture bounding box.north west)
		 (path picture bounding box.south west) -- (path picture bounding box.north east);}},
	   zigzag/.style = {
	   	to path={ -- ($(\tikztostart)!.55!-9:(\tikztotarget)$) --
	   		($(\tikztostart)!.45!+9:(\tikztotarget)$) -- (\tikztotarget)
	   		\tikztonodes},sharp corners},
  multiple/.style = 	{double copy shadow={shadow xshift=0.25em,shadow yshift=-0.25em,draw=black!30, left color=lightgreen!50!white, right color=lightgreen!10!white},fill=white,draw=black,thick,minimum height = 5.5em,minimum width=9.5em, rounded corners=0.1cm,left color=lightgreen!50!white, right color=lightgreen!10!white},
   		}
\tikzstyle{branch}=[{circle,inner sep=0pt,minimum size=0.3em,fill=black}]
\tikzstyle{box} = [draw, dotted, inner xsep=4mm, inner ysep=3mm]
\tikzstyle{every path}=[line width=0.1em]
\begin{tikzpicture}[auto, line width=0.1em, node distance = 1em]

\node[multiple, align=center](shadow){};
\node[block, at={(shadow.north)}, fill=none, yshift=-1.5em](ms){Minimum Statistics};
\node[block, below= of ms,fill=none](wf){Wiener filter};
\node[block, below= 1.5em of shadow, bottom color=yellow!50!white, top color=yellow!10!white](hp){Highpass};
\node[block, below= of hp, bottom color=yellow!50!white, top color=yellow!10!white](lpc){LPC};
\node[block, right=3em of wf, yshift=2em, top color=red!40!white, bottom color=red!15!white](et){Energy Tracker};
\node[block, below= of et.south east, anchor=north east, minimum width=4em, top color=red!40!white, bottom color=red!15!white](em){EM-Alg.};
\node[block, below=4em of et, top color=red!40!white, bottom color=red!10!white](viterbi){HMM/Viterbi};

\draw[-latex] (ms) -- (wf);
\draw[-latex] ($(shadow.south) - (0,0.5em)$) -- (hp);
\draw[-latex] (hp) -- (lpc);
\draw[-latex] (lpc.south) -- ($(lpc.south) - (0,0.5em)$) -| ($(wf) !0.5! (et)$) |- ($(et.north) + (0,1em)$) -- (et.north);
\draw[-latex] ($(et.south) - (2.5em,0)$) -- ($(viterbi.north) - (2.5em,0)$);
\draw[-latex] ($(et.south) - (2.5em,0)$) |- (em.west);
\draw[-latex] (em.south) -- (em.south |- viterbi.north);
\draw[-latex] ($(shadow.north) + (0,1.5em)$) -- node[above, pos=0.1]{Audio} (shadow.north);
\draw[-latex] (viterbi.south) -- node[below, pos=0.9]{SAD} ($(viterbi.south) - (0,1.5em)$);
\end{tikzpicture}
	\caption{Overview on statistical \gls{SAD} components and signal processing queue.}
	\label{fig:sigpro}
        \vspace{-0.5cm}
\end{figure}

Our approach is depicted in \cref{fig:sigpro}: It consists of repeated application of a denoising stage (\cref{fig:sigpro}, green blocks), high-pass and \gls{LPC} filtering (\cref{fig:sigpro}, yellow blocks), and the statistical \gls{SAD} (\cref{fig:sigpro}, orange blocks).
Each block is described in the following.


As input to the statistical \gls{SAD} we choose the \gls{STFT}-coefficients of the observed signal $X_{t,f}$ with $t\in[0,T]$ as the frame index and $f\in[0,F]$ as the frequency bin index.
From these coefficients, the minimum statistics based noise \gls{PSD} estimate $\overline{|V(t,f)|^{2}}$, and $|X(t,f)|^{2}$, i.e., the \gls{PSD} estimate of the current analysis window, are calculated and the corresponding Wiener filter $W(t,f)$ is given by:
\begin{align}
W(t,f) = \mathrm{max}\left(1-\gamma\frac{\overline{|V(t,f)|^{2}}}{|X(t,f)|^{2}}, G_{\mathrm{min}}\right)
\end{align}
The oversubtraction factor $\gamma$ is chosen relatively high, $\gamma > 20$, to compensate for the bias of underestimating the noise level via minimum statistics and to force the Wiener Filter to aggressively apply noise reduction.
Furthermore, $W(t,f)$ is lower bounded by $G_\mathrm{min}$ to
prevent $W(t,f)$ from becoming negative in case of low noise levels co-occuring with speech absence in the same bin.
Noise tracking and Wiener filtering iterate multiple times over the audio signal, decreasing the noise level in each iteration and at the same time keeping the maximum peaks corresponding to speech untouched. Thereby, the \gls{SNR} is improved with each stage, however, at the cost of deteriorating the audio quality. Since neither a low number of acoustic artifacts (e.g., musical tones), nor superior speech quality are of interest here, this loss is acceptable.

\begin{figure}[t]
	\centering

%
%
\begin{tikzpicture}

\begin{axis}[%
width=0.4\columnwidth,
height=2cm,
at={(1.5in,0in)},
scale only axis,
axis on top,
xmin=-0.016,
xmax=79.952,
x label style={yshift=0.6em},
xlabel={Time [s]},
ymin=-0.0078125,
ymax=4.0078125,
legend style={legend cell align=left,align=left,draw=white!15!black}
]
\addplot [forget plot] graphics [xmin=-0.016,xmax=79.952,ymin=-0.0078125,ymax=4.0078125] {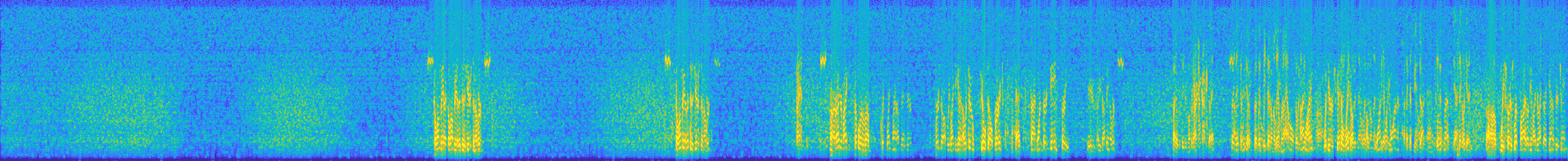};
\end{axis}

\begin{axis}[%
width=0.4\columnwidth,
height=2cm,
at={(0in,0in)},
scale only axis,
axis on top,
xmin=-0.016,
xmax=79.952,
x label style={yshift=0.6em},
xlabel={Time [s]},
ymin=-0.0078125,
ymax=4.0078125,
y label style={yshift=-1.8em},
ylabel={Frequency [kHz]},
legend style={legend cell align=left,align=left,draw=white!15!black}
]
\addplot [forget plot] graphics [xmin=-0.016,xmax=79.952,ymin=-0.0078125,ymax=4.0078125] {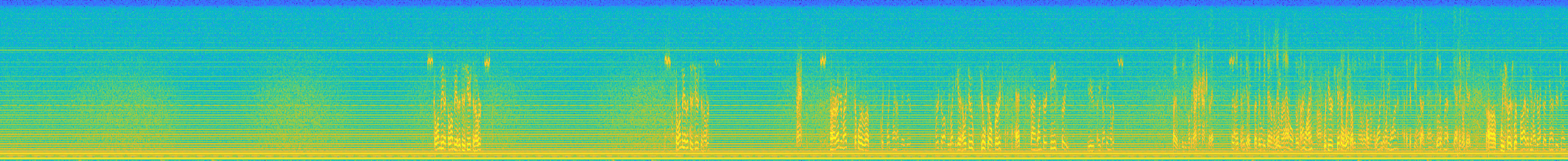};
\end{axis}
\end{tikzpicture}%
	\vspace{-0.3cm}
	\caption{Input signal (left) and processed signal (right).}
        \vspace{-0.5cm}
\end{figure}

After the denoising stage, a linear highpass filter is applied to remove low frequency noise. Furthermore, a simple $1^{\text{st}}$-order \gls{LPC} filter is employed to enhance the well predictable speech signals and to suppress the unpredictable noise.

From the enhanced audio signal the energy per sub-band is calculated, where each sub-band has a bandwidth of $\SI{1}{kHz}$. Temporal smoothing with an averaging window of size $\SI{0.48}{s}$ reduces the sub-band energy variance. Afterwards, the smoothed sub-band energies are weighted with an exponential decay factor ($1/s$ for the $s^{\text{th}}$ sub-band) and accumulated to a single value per frame, called \gls{CSBE}$(t)$.

\subsection{Adaptive threshold}
The \gls{CSBE} values are the basic information source for deciding whether an active speaker or just noise is observed.
However, finding an optimal decision threshold is a nontrivial task. Furthermore, the resulting threshold may be dependent on the dataset. 
In order to circumvent a fixed threshold, minimum statistics is applied once more. This time, the \gls{CSBE} values are tracked to find the floor values (F-CSBE), which belong to the non-speech parts of the recordings. Additionally, the mean of all F-CSBE values for each recording can be calculated to get the average \gls{CSBE} floor value (A-CSBE), i.e., an estimate for the average noise level of the recording. A frame is marked as speech if \gls{CSBE} exceeds the sum of F-CSBE and A-CSBE by a certain factor.

\begin{figure}[b]
        \vspace{-0.25cm}
	\centering
	\input{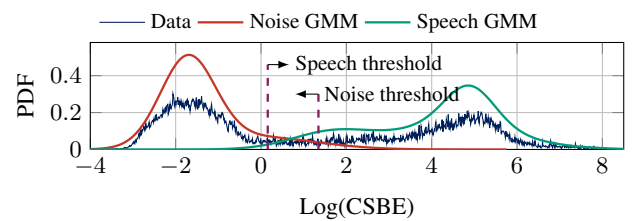}
	\vspace{-0.3cm}
	\caption{Histogram of logarithmic CSBE values with noise and speech GMM models.}
	\label{Fig:models2}
\end{figure}

The described minimum tracking approach delivers an initial estimate of speech activity, but it shows a high sensitivity towards the chosen threshold.
To overcome this issue, a statistical model in the log-domain is established.
As depicted in \cref{Fig:models2}, the logarithm of the \gls{CSBE} values is taken and all values smaller than the noise threshold are considered to be caused by noise and thus used to estimate a \gls{GMM} representing noise components.
Similarly, values larger than the chosen speech threshold are considered  speech  and used to estimate a \gls{GMM}  for speech.
The thresholds for speech and noise are derived from the A-CSBE value, adding some additional safety margins.

The final decision of the statistical \gls{SAD} is derived with a Viterbi decoder operating on an \gls{HMM} which consists of $5$ consecutive states for noise and $5$ states for speech, each with probability of $0.9$ for staying in the state and $0.1$ for state switching.
The \gls{HMM} emission probabilities are given by the aforementioned \glspl{GMM}.

\section{Neural Network-Based SAD}
\label{sec:dnn}
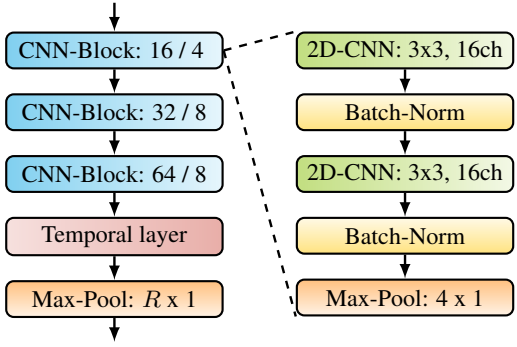
\begin{figure}[t]
	\centering
	\tikzset{%
  block/.style    = {draw, thick, rectangle, minimum height = 1.5em, minimum width = 9em, fill=white, align=center, rounded corners=0.1cm},
  sum/.style      = {draw, circle, node distance = 2cm}, 
  cross/.style={path picture={\draw[black](path picture bounding box.south east) -- (path picture bounding box.north west)
		 (path picture bounding box.south west) -- (path picture bounding box.north east);}},
	   zigzag/.style = {
	   	to path={ -- ($(\tikztostart)!.55!-9:(\tikztotarget)$) --
	   		($(\tikztostart)!.45!+9:(\tikztotarget)$) -- (\tikztotarget)
	   		\tikztonodes},sharp corners}
               }
\tikzstyle{branch}=[{circle,inner sep=0pt,minimum size=0.3em,fill=black}]
\tikzstyle{box} = [draw, dotted, inner xsep=4mm, inner ysep=3mm]
\tikzstyle{every path}=[line width=0.1em]
\begin{tikzpicture}[auto, line width=0.1em]
\node[block, left color=lightblue!50!white, right color=lightblue!10!white, at={(0,0)}] (cnn2d) {CNN-Block: 16 / 4 } ;
\node[block, left color=lightblue!50!white, right color=lightblue!10!white, below= 1em of cnn2d] (cnn2d_2) {CNN-Block: 32 / 8} ;
\node[block, left color=lightblue!50!white, right color=lightblue!10!white, below= 1em of cnn2d_2] (cnn2d_3) {CNN-Block: 64 / 8} ;
\node[block, below= 1em of cnn2d_3, right color=red!40!white, left color=red!15!white] (smoothing) {Temporal layer} ;
\node[block, below = 1em of smoothing, top color=orange!50!white, bottom color=orange!10!white] (maxpool) {Max-Pool: $R$ x 1};

\node[block, right= 3em of cnn2d, left color=lightgreen!50!white, right color=lightgreen!10!white] (cnn2d_exp_1) {2D-CNN: 3x3, 16ch} ;
\node[block, below= 1em of cnn2d_exp_1, bottom color=yellow!50!white, top color=yellow!10!white] (batch_norm) {Batch-Norm} ;
\node[block, below= 1em of batch_norm, left color=lightgreen!50!white, right color=lightgreen!10!white] (cnn2d_exp_2) {2D-CNN: 3x3, 16ch} ;
\node[block, below= 1em of cnn2d_exp_2, bottom color=yellow!50!white, top color=yellow!10!white] (batch_norm_2) {Batch-Norm} ;
\node[block, below = 1em of batch_norm_2, top color=orange!50!white, bottom color=orange!10!white] (maxpool_exp) {Max-Pool: 4 x 1};

\draw[-latex] (0, 2em) -- (cnn2d);
\draw[-, dashed] (cnn2d.east) -- (cnn2d_exp_1.north west);
\draw[-, dashed] (cnn2d.east) -- (maxpool_exp.south west);
\draw[-latex] (cnn2d) -- (cnn2d_2);
\draw[-latex] (cnn2d_2) -- (cnn2d_3);
\draw[-latex] (cnn2d_3) -- (smoothing);
\draw[-latex] (smoothing) -- (maxpool);
\draw[-latex] (cnn2d_exp_1) -- (batch_norm);
\draw[-latex] (batch_norm) -- (cnn2d_exp_2);
\draw[-latex] (cnn2d_exp_2) -- (batch_norm_2);
\draw[-latex] (batch_norm_2) -- (maxpool_exp);
\draw[-latex] (maxpool.south) -- ($(maxpool.south) - (0,1em)$);

\end{tikzpicture}
	\caption{Block diagram of the  \gls{DNN} model used for \gls{SAD}, where $R$ symbolizes the output size of the temporal layer.}
	\label{fig:model}
        \vspace{-0.25cm}
\end{figure}

The neural network architecture is adapted from \cite{Ebbers2019WeaklySAD} and consists of multiple \gls{CNN} blocks with subsequent layers for temporal smoothing, an output activation and max pooling over the feature dimension.
Each \gls{CNN} block consist of two 2D-\gls{CNN} layers with a $3$x$3$ kernel and a subsequent batch normalization, followed by a single
max pooling layer of stride 4 which operates along the feature dimension.
In contrast to \cite{Ebbers2019WeaklySAD}, no pooling along the time dimension is applied to ensure frame-wise outputs.
To exploit temporal information, we either use a 1D-\gls{CNN} structure as in \cite{Ebbers2019WeaklySAD} or an \gls{RNN} layer with a bidirectional \gls{GRU} \cite{Kyunghyun14GRU} and a subsequent \gls{FF} layer as classifier to enable the network to use a larger context.
The described network structure is shown in \cref{fig:model} and  \cref{tbl:temporal_layer} compares the two options for gathering temporal information.

\begin{table}[b]
        \vspace{-0.25cm}
	\caption{%
		Comparison of two possible layer structures for the temporal layer where FF represents a feed forward layer.
	}
	\label{tbl:temporal_layer}
	\renewcommand*{\arraystretch}{1.1}
	\centering
	\footnotesize
	\begin{tabular}{lccc}
		\toprule
		Layer type & $\#$Layer & Params & Classifier\\
		\toprule
		1D-CNN & 2 & 3x1 Kernel / (128,10)ch & --\\
		RNN & 1 & BI-GRU: 64x256  & FF: 256x10 \\
		\toprule
	\end{tabular}
\end{table}

\begin{figure}[t]
	\centering
	\tikzset{%
	block/.style    = {draw, thick, rectangle, minimum height = 1.5em, minimum width = 3.2em, fill=white, align=center, top color=blue!40!white, bottom color=blue!10!white},
	sum/.style      = {draw, circle, node distance = 2cm}, 
	cross/.style={path picture={\draw[black](path picture bounding box.south east) -- (path picture bounding box.north west)
			(path picture bounding box.south west) -- (path picture bounding box.north east);}},
	box/.style={draw, dashed, inner xsep=0mm, inner ysep=0mm, line width=0.1em}
}
\begin{tikzpicture}[line width=0.1em]
\begin{scope}[node distance=0em and 0em]
\node[block](x1){$x_1$};
\node[block, right= of x1](x2){$x_2$};
\node[block, right= of x2](x){$...$};
\node[block, right= of x](xz){$...$};
\node[block, right= of xz](xN1){$x_{N\mathrm{-}1}$};
\node[block, right= of xN1](xN){$x_N$};

\node[block, below= 1em of x1](yx1){$x_1$};
\node[block, right= of yx1](yx2){$...$};
\node[block, right= of yx2](yx3){$x_L$};
\node (box) [box, red!80!white, fit=(yx1.north west) (yx3.south east)] {};

\node[block, below= of yx2, xshift=1.6em](yy1){$...$};
\node[block, right= of yy1](yy2){$...$};
\node[block, right= of yy2](yy3){$...$};
\node (box) [box, red!80!white, densely dotted, fit=(yy1.north west) (yy3.south east)] {};

\node[block, below= of yy2, xshift=1.6em](yz1){$x_M$};
\node[block, right = of yz1](yz2){$...$};
\node[block, right = of yz2](yz3){$x_N$};
\node (box) [box, red!80!white, dashdotted, fit=(yz1.north west) (yz3.south east)] {};

\draw  (yx1.west |- yy1.west) edge[latex-latex] node[below]{$S$}(yy1.west);

\node[block, below = 1em of yz1, right color=red!40!white, left color=red!15!white, rounded corners=0.1cm] (gru3) {RNN};
\node[block, at={(gru3 -| yy1)}, right color=red!40!white, left color=red!15!white, rounded corners=0.1cm] (gru2) {RNN};
\node[block, at={(gru3 -| yx1)}, right color=red!40!white, left color=red!15!white, rounded corners=0.1cm] (gru1) {RNN};

\node[block, below= 1em of gru2](out1){$y_1$};
\node[block, right = of out1](out){$...$};
\node[block, right= of out](out2){$y_M$};
\coordinate (t) at ($(out.south)  - (0, 0.5em)$);
\draw[dashed, line width=0.2em, lightgreen] (x1.west |- t) -- node[above, pos=0.9, black](writting){apply} node[below, pos=0.9, black]{threshold} (xN.east |- t);

\node[block, below= 1em of out1, xshift=-1.6em](z1){$...$};
\node[block, below= of z1, xshift=1.6em](z2){$...$};
\node[block, left= of z1](zx1){$d_1$};
\node[block, right = of z2](zx2){$...$};
\node[block, right= of zx2](z3){$...$};
\node[block, below= of z3, xshift=-1.6em](zx3){$d_M$};
\node[block, right= of z1](zz1){$d_1$};
\node[block, right= of zx3](zz2){$...$};
\node[block, right= of zz2](zz3){$d_M$};
\node[block, below =1em of zz3](o2){$v_N$};
\node[block, left= of o2](o1){$v_{N-1}$};
\node[block, left= of o1](o3o){$..$};
\node[block, left= of o3o](o3){$..$};
\node[block, left= of o3](o4){$v_2$};
\node[block, left= of o4](o5){$v_1$};
\end{scope}

\draw[-latex] (x.south -| x.east) -- (yx3.north -| yx3.east);
\draw[-latex, red!80!white, dashed] (yx1.south) -- (gru1.north);
\draw[-latex, red!80!white, densely dotted] (yy1.south) -- (gru2.north);
\draw[-latex, red!80!white, dashdotted] (yz1.south) -- (gru3.north);
\draw[-latex] (gru1.south) -- (out1.north);
\draw[-latex] (gru2.south) -- (out.north);
\draw[-latex] (gru3.south) -- (out2.north);
\node[draw=none, fill=none, at={(gru2-|writting)}, align=center] (params){shared\\parameters};
\coordinate (zw) at ($(gru1.west) - (0.5em, 0)$);
\node (box) [box, orange, fit=(zw) (params.north east) (params.south east)] {};
\draw[-latex] (out) -- (out.south |- zz1.north);
\draw[-latex] (zx2.south |- zz3.south) -- (zx2.south |- o3.north);
\draw[decorate,decoration={brace,amplitude=1em}] (zx1.west |- zx3.south) -- (zx1.north west) node[midway, rotate=90, yshift=0.5cm]{max};
\draw  (zx1.west |- z2.west) edge[latex-latex] node[below]{$S$}(z2.west);
\end{tikzpicture}
	\caption{Example of the segment \gls{RNN}}
	\label{fig:segmented_gru}
\vspace{-0.7cm}
\end{figure}
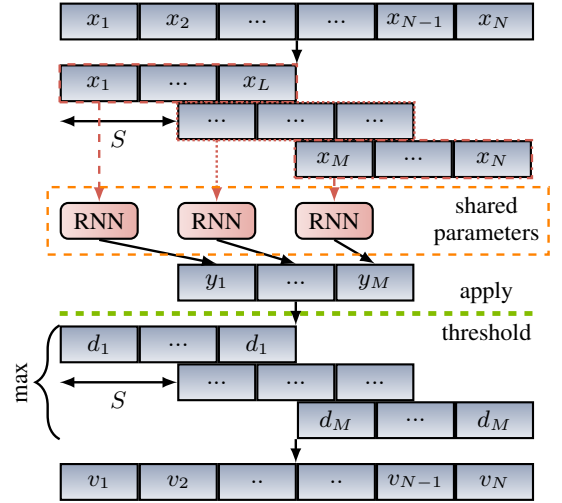

To enforce temporal smoothing, the common approach is to use Viterbi decoding on an \gls{HMM} with \glspl{GMM} emission distributions to model speech and noise statistics, as described in the  previous section.
However, the smoothing can also be done by the \gls{DNN} or, more specifically, by the temporal layer.
Therefore, we propose a segment \gls{RNN} as replacement for the temporal layer where each input utterance is segmented into $M$ overlapping chunks of length $L$ with a shift $S$ between segments as outlined in \cref{fig:segmented_gru}.
Each segment is processed by an \gls{RNN} with subsequent classifier layer as specified in \cref{tbl:temporal_layer}.
The parameters are shared between the layers to ensure that all segments are processed equally. 
For each segment only the last output frame is chosen as the prediction for the whole segment.
Speech activity is assumed for a segment $i$ if the estimated output $y_i$ exceeds a certain threshold $\alpha$:
$
d_i = \begin{cases}
1 \text{ if } y_i > \alpha\\
0 \text{ if } y_i \leq \alpha
\end{cases}$.
Finally, speech presence is declared for a given frame if at least one segment containing that frame indicates speech presence.
Thereby, the occurrence of oscillations in the decision signal is reduced at the cost of overestimating the speech activity.
Since the segment length $L$ and shift $S$ are fixed, the segment \gls{RNN} approach may lead to a higher hit rate while also increasing the false alarm rate.
However, in most applications a high \gls{TPR} is more important than a low \gls{FPR}. For example, in case of \gls{ASR} an overestimation of the length of speech activity is not as harmful as missing part of an utterance.
The shift $S$ allows for a complexity reduction since a larger value reduces the number of segments and thus the number of chunks to be processed by the \gls{RNN}.
However, increasing $S$ also reduces the overlap between segments and thereby the number of segments contributing to the activity estimation for each frame.
Changes in the segment length $L$, control the temporal context seen by the \gls{RNN} layer.

For training, the \SI{30}{min} streams are divided randomly into \SI{4}{s} segments which are independently sent through the network,
This prevents overfitting since it ensures that all possible speech-silence-ratios are observed during training.
As cost function the binary cross entropy is chosen.


\section{Evaluation}
The presented \gls{SAD} systems are tuned on the development set of the fearless dataset \cite{Hansen18Fearless}. As metrics we use precision ($P$), recall ($R$), $F_1$-score$=2\frac{P\cdot R}{P+R}$ and \gls{DCF} with\\
\begin{minipage}{0.5\columnwidth}
	\begin{align}
		P &=\frac{\mathrm{TP}}{\mathrm{TP} + \mathrm{FP}},
	\end{align}
\end{minipage}
\begin{minipage}{0.49\columnwidth}
	\begin{align}
		R &= \frac{\mathrm{TP}}{\mathrm{TP} + \mathrm{FN}},
	\end{align}
\end{minipage}
\begin{align}
\mathrm{DCF} &= 0.75 \cdot \frac{\mathrm{FN}}{\mathrm{TP} + \mathrm{FN}} + 0.25\cdot \frac{\mathrm{FP}}{\mathrm{TN} + \mathrm{FP}},
\end{align}
where TP, FP, TN and FN are the number of true positive, false positive, true negative and false negative predictions.
The scoring is done with the openSAD evaluation tool \cite{nist16}.

\subsection{Fearless Steps Dataset}
The Fearless Steps dataset \cite{Hansen18Fearless} consists of $\SI{8}{kHz}$ recordings from the Apollo 11 mission.
The part of the Fearless Steps dataset used during this challenge for training and development consists of 290 speakers with an \gls{SNR} between \SI{0}{dB} and \SI{20}{dB}.
Note, that the development set includes 34 unique speakers not seen during training.
All examples are \SI{30}{min} long, and the training set consists of $\SI{29.56}{\percent}$ of speech on average, whereas the development set includes an average of $\SI{32.87}{\percent}$ of speech.
The main challenges of the dataset are, first, that speech activity is typically very short, consisting of one or two words only, and, second, that the noise is highly non-stationary  with varying \gls{SNR}.
If not stated otherwise, all experiments are carried out on the development set.


\subsection{DNN tuning}
\begin{figure}[t]
	\centering
	\input{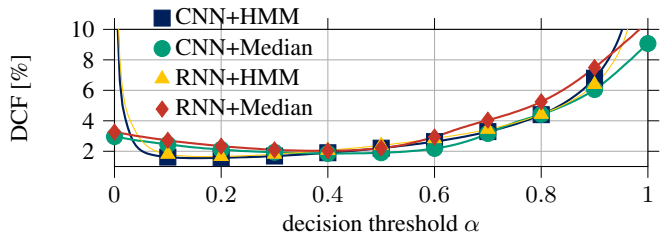}
	\vspace{-0.7cm}
	\caption{Comparison of the \gls{CNN} and \gls{RNN} temporal layer with median filter or \gls{HMM}-based smoothing}
	\label{fig:cnn_rnn}
\vspace{-0.5cm}
\end{figure}
The input to the \gls{DNN} is the magnitude spectrum obtained from an \gls{STFT}  with an FFT size of $512$ samples, a window length of \SI{50}{ms} and a frame shift of \SI{10}{ms}.
All networks were trained for $50000$ iterations with a batch size of $24$ and a learning rate of $0.001$ using the Adam optimizer \cite{Kingma2014AdamAM}.

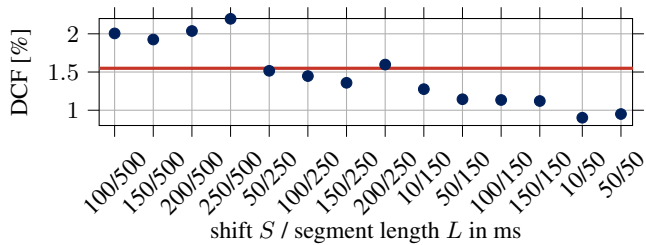
\begin{figure}[b]
	\centering
\begin{tikzpicture}

\begin{axis}[
height=3cm,
tick align=outside,
tick pos=both,
width=\columnwidth,
x grid style={white!69.01960784313725!black},
xlabel={shift $S$ / segment length $L$ in ms},
xmajorgrids,
xmin=-0.4, xmax=13.4,
xtick style={color=black},
xtick={0,1,2,3,4,5,6,7,8,9,10,11,12.1,13.1},
xticklabel style = {rotate=45.0},
xticklabels={100/500,150/500,200/500,250/500,50/250,100/250,150/250,200/250,10/150,50/150,100/150,150/150,10/50,50/50},
y grid style={white!69.01960784313725!black},
ylabel={DCF [\%]},
xlabel style = {yshift=-2em},
ymajorgrids,
ymin=0.8, ymax=2.2,
ytick style={color=black},
ylabel near ticks
]
\addplot [only marks, mark=*, draw=blue, fill=blue, colormap/viridis]
table{%
x                      y
0 2.0051767619224257
1 1.9251955607308323
2 2.036593799497593
3 2.1961361139570288
4 1.5158808279580754
5 1.4469650457115565
6 1.358876802335931
7 1.5974351270769665
8 1.275831901519005
9 1.1437756683437335
10 1.133945342148483
11 1.1206170843490647
12.1 0.901939190429262
13.1 0.9501894915145309
};
\addplot [very thick, red]
table {%
-1 1.54887430998224
15 1.54887430998224
};
\end{axis}

\end{tikzpicture}
	\vspace{-0.7cm}
	\caption{DCF results for the segment \gls{RNN} with different shifts $S$ and segment length $L$ for the respective optimal threshold. The red line symbolizes the DCF value for the CNN+HMM system.}
	\label{fig:shift_length}
\vspace{-0.25cm}
\end{figure}

In \cref{fig:cnn_rnn}, the \gls{DCF} is displayed as a function of the decision threshold $\alpha$ for the first two temporal layer variants described in \cref{sec:dnn}.
According to the figure, \gls{RNN} and the 1D-\gls{CNN} achieve similar results with a small edge for the 1D-\gls{CNN}.
In both cases, the threshold can be chosen in a fairly large range without a substantial impact on the performance.

Additionally, two types of temporal smoothing are compared:
On one hand, \gls{HMM}-based smoothing described in \cref{sec:stat_sad} and, on the other hand, a median filter with a fixed window length.
It can be observed that the \gls{HMM}-based approach outperforms the median filter for both the \gls{RNN} and \gls{CNN} layer.

In \cref{fig:shift_length}, the \gls{DCF} values achieved with the segment \gls{RNN} as temporal layer are plotted for different shifts $S$ and lengths $L$.
For each $S$ and $L$ the individual optimal threshold $\alpha$ is chosen.
The results are compared to the previously best presented system, the \gls{CNN}+\gls{HMM}.
It is apparent that the segment \gls{RNN} clearly outperforms the \gls{CNN}+\gls{HMM} \gls{SAD} estimation for short segment length $L$.
All results with $L\leq \SI{250}{ms}$ and $S\leq \SI{150}{ms}$ achieve at least a small gain compared to the \gls{CNN} temporal layer with \gls{HMM} smoothing.
Reducing the shift and length in the segment \gls{RNN} to $\SI{10}{ms}$ and $\SI{50}{ms}$ which equals $1$ and $5$ frames respectively, results in the lowest \gls{DCF}. 
A possible explanation is the non-stationarity of the distortions 
and that the high overlap between segments due to the small shift successfully smooths the network output.
%


\subsection{System comparison}
In \cref{tbl:comparison} the results for all systems tuned to their optimal threshold are shown.
The presented statistical \gls{SAD} achieves a \gls{DCF} of \SI{2.98}{\percent} and thus an improvement of \SI{9.52}{\percent} over the challenge baseline in terms of absolute numbers.
In comparison, the neural network-based approaches achieve a \gls{DCF} of \SI{1.42}{\percent} and \SI{1.54}{\percent} with a \gls{CNN} and \gls{RNN} as temporal layer, respectively.
However, the segment \gls{RNN} temporal layer outperforms all other approaches 
achieving a \gls{DCF} of \SI{0.81}{\percent}.
The high precision $P$ indicates that a small segment length $L$ and shift $S$ allows to increase the hit rate without causing an higher false alarm rate.
The results achieved on the evaluation dataset are similar to the ones observed on the development data for both the \gls{DNN}-based and statistical \gls{SAD}.
Indicating, that the systems are robust to small changes in the data.
Using majority voting on the output of all neural networks in \cref{tbl:comparison} we get a \gls{DCF} of \SI{1.07}{\%} which is the best submitted \gls{SAD} result during the 2020 Fearless Steps Challenge but only a slight improvement over the single model with a segment \gls{RNN} temporal layer. 

\begin{table}[t]
	\caption{%
		Results for all presented systems for different metrics in $\%$ on the Dev and Eval set for the Fearless Steps SAD Challenge.
	}
	\label{tbl:comparison}
	\renewcommand*{\arraystretch}{1.1}
	\centering
	\footnotesize
	\begin{tabular}{
			l
			S[table-auto-round, table-format=-1.2]
			S[table-auto-round, table-format=-1.2]
			S[table-auto-round, table-format=-1.2]
			S[table-auto-round, table-format=-1.2]
			S[table-auto-round, table-format=-1.2]
		}
		\toprule
		& \multicolumn{4}{c}{DEV} & {EVAL}\\
		System & {F1} & {$P$} & {$R$} & {DCF} & {DCF} \\
		\toprule
		Stat. SAD & 94.32217526750351 & 90.93731660445404 & 97.96875743753867 & 2.9834474309412573 & 4.597 \\
		CNN + HMM & 97.91381610240492 & 96.82945514422941 & 99.02273891824238 & 1.417806020104808 & 2.453 \\
		RNN + HMM & 97.8434915637541 & 96.85487521501645 & 98.85249803298977 & 1.540646372455483 & 2.173 \\
		Seg. RNN & 98.62083174888512 & 97.42280771731589 & 99.8486872131735 & 0.8084724390954476 & 1.187 \\
		Baseline & {--} & {--} & {--} & 12.5 & 13.6 \\
		\toprule
	\end{tabular}
\vspace{-0.65cm}
\end{table}

The two proposed approaches differ in many aspects, e.g. their type of signal processing or their implementations. In \cref{tbl:complexity} a comparison between the systems in terms of processing time is stated.
Please note that the table shows results for systems which are implemented on different tool chains, and that further optimizations on the code may improve the realtime factors.
The table shows that all systems allow real time processing.

\begin{table}[b]
        \vspace{-0.5cm}
	\caption{%
		Realtime factor of \gls{SAD} on an
		[Intel\textsuperscript \textregistered Xeon\textsuperscript \textregistered CPU E3-1240 v6 @ 3.70GHz, 8GB RAM].
	}
	\label{tbl:complexity}
	\newcommand{\NUM}[1]{\num[round-precision=3,round-mode=figures]{#1}}
	\centering
	\footnotesize
	\begin{tabular}{lccc}
		\toprule
		System & {Stat. SAD} & {CNN + HMM} & {Seg. RNN} \\
		\toprule
		Realtime factor & \NUM{0.0047198272} & \NUM{0.01193751154402215} & \NUM{0.044466430436588496}\\
		Tool chain & C++/Matlab & Pytorch \cite{pytorch19} & Pytorch \cite{pytorch19}\\
		\toprule
	\end{tabular}
\end{table}

Although the statistical \gls{SAD} achieves worse detection rates than the \gls{DNN}-based method presented here, the approach is interesting because it has a lower real-time factor and because it is an unsupervised learning approach not requiring labeled training data. Arguably, this makes it easier to adapt the system to other data sets.






\section{Conclusions}
This paper proposes a new statistical \gls{SAD} which achieves competitive results compared to other \gls{DNN}-based systems.
Furthermore, a new \gls{DNN}-based \gls{SAD} with a segment \gls{RNN}-based smoothing is presented which allows to define the context observed by the \gls{RNN}.
This single system approach achieves a DCF value of \SI{1.19}{\%} which is the second best result submitted to the 2020 Fearless Steps Challenge and is only slightly outperformed by a majority voting between a combination of the proposed \gls{DNN} architectures (\gls{DCF}: \SI{1.07}{\%}).
Hereby, the majority voting results are the best submitted to the 2020 Fearless Steps Challenge, outperforming the baseline by  more than \SI{10}{\percent} in absolute values.

\section{Acknowledgement}
The authors would like to thank Plath GmbH in Hamburg (Germany) for funding part of the research for this contribution.
\balance
\bibliographystyle{IEEEtran}
\bibliography{mybib}

\end{document}